# Local action of curl-less vector potential on vortex dynamics in superconductors

Ellen D. Gulian, Gurgen G. Melkonyan, Vahan R. Nikoghosyan, and Armen M. Gulian[1]

*Advanced Physics Laboratory, Institute for Quantum Studies, Chapman University, 15202 Dino Dr., Burtonsville, MD 20866, USA*

*Abstract.* - Study of the Abrikosov vortex motion in superconductors based on time-dependent Ginzburg-Landau equations reveals an opportunity to locally detect the values of the Aharonov-Bohm type curl-less vector potentials.

> *"**E** and **B** are slowly disappearing from the modern expression of physical laws; they are being replaced by **A** and $\phi$."*
>
> R. Feynman [1]

## 1. Introduction

Ever since Heaviside and Hertz eliminated the need for the vector potential ***A*** in Maxwell's equations, it was believed that ***A*** did not fit the definition of a real field and that it was simply a mathematical tool [2]. However, phenomena involving superconductors, such as the Aharonov-Bohm (AB) effect [3] in which electrons diffract in the presence of the curl-less ***A***-field [4], provide proof that ***A*** is in fact a "real" field. In his classical treatise [1], R. Feynman defines a real field as a "mathematical device for avoiding the idea of action at a distance." Commenting on the AB-effect, he comes to the conclusion that "the classical electromagnetic field acting locally on a particle is not sufficient to predict its quantum-mechanical behavior" and that the vector potential is a real field. In this article, we will provide an example based on solutions of time-dependent Ginzburg-Landau (TDGL) theory where the curl-less ***A***-field creates classically detectable effects in quasi-local quantum objects (such as short thin superconducting strips) in a dynamic regime.

## 2. Model

Consider the motion of current in a thin superconducting film of restricted geometry with two symmetric dents, as shown in Fig. 1.

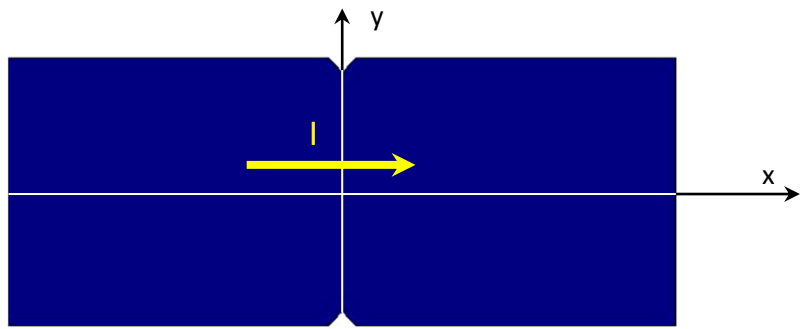

Fig. 1. A thin current-carrying superconducting film (20x8 in $\lambda_L$ units) with two symmetrically arranged dents. Total current through the film cross section is I. Film thickness is less than $\lambda_L$; the physical picture is therefore dependent on x- and y-coordinates only.

---

[1] Corresponding author. E-mail: gulian@chapman.edu

We will assume (and justify through later calculations) that the physical pattern is symmetric relative to the x-axis.

## 3. Basic equations

To describe the current flow through this film, we utilize a set of TDGL equations [5]. For simplicity, we will not include the non-equilibrium phonon term, and the equation for the order parameter $\Delta = |\Delta|\exp(i\theta)$ will have the form:

$$-\frac{\pi}{8T_c}\frac{1}{\sqrt{1+(2\tau_\varepsilon|\Delta|)^2}}\left(\frac{\partial}{\partial t}+2i\varphi+2\tau_\varepsilon^2\frac{\partial|\Delta|^2}{\partial t}\right)\Delta +$$
$$+\frac{\pi}{8T_c}\left[D(\nabla-2iA)^2\right]\Delta+\left(\frac{T_c-T}{T_c}-7\zeta(3)\frac{|\Delta|^2}{8(\pi T_c)^2}\right)\Delta = 0. \tag{1}$$

Here, $A$ and $\varphi$ are vector and scalar potentials of the electromagnetic field, $T_c$ is the critical temperature of the superconductor, $\tau_\varepsilon$ is the electron-phonon relaxation time, $D$ is the electronic diffusion coefficient, and $\zeta(3)$ is the Riemann zeta function. In these expressions, the theoretical units $\hbar = c = e = 1$ are used. For the numerical modeling which follows, Eq. (1) should be rewritten in its dimensionless form. Dividing Eq. (1) by $(T_c - T)/T_c$ and introducing the coherence length $\xi(T)$

$$\xi^2(T) = \frac{\pi D}{8(T_c - T)}, \tag{2}$$

as well as the equilibrium temperature-dependent value of the order parameter $\Delta_0(T)$:

$$\Delta_0^2 = \frac{8\pi^2 T_c}{7\zeta(3)}(T_c - T). \tag{3}$$

and denoting $\Psi = \Delta/\Delta_0$, $\Gamma = 2\tau_\varepsilon \Delta_0$, $\tau = tD/\xi(T)^2$ and $\phi = 2\varphi\xi(T)^2/D$, we obtain the equation:

$$\frac{1}{\sqrt{1+\Gamma^2|\Psi|^2}}\left(\frac{\partial}{\partial\tau}+i\phi+\frac{1}{2}\Gamma^2\frac{\partial|\Psi|^2}{\partial\tau}\right)\Psi = -\left(\frac{i}{k}\nabla+\mathrm{A}\right)^2\Psi+\left(1-|\Psi|^2\right)\Psi. \tag{4}$$

Here $\kappa = \lambda_L/\xi$ is the Ginzburg-Landau parameter, and the vector potential is renormalized as $\mathrm{A} = 2A\xi$ (for notational purposes, we do not use bolded letters for this quantity hereafter).

  In the gapless regime ($\Gamma = 0$) this equation coincides with those used by other researchers (see, *e.g.*, [6,7]). We note in passing that the application of TDGL equations does not require one to be in the gapless regime. Rather, these equations are valid for a much wider range of "finite-gap" superconductors.

The second equation required to complete the system is the equation for the total current, which we will write as

$$\sigma\left(\frac{\partial A}{\partial t} + \nabla\phi\right) = \frac{1}{2ik}\left(\Psi^*\nabla\Psi - \Psi\nabla\Psi^*\right) - |\Psi|^2 A - \nabla\times\nabla\times A. \quad (5)$$

Note that these equations (4) and (5) are gauge-invariant [5], and we can use the most convenient gauge for our modeling. We chose the gauge $\phi = 0$, so that the final set of equations is:

$$\dot{\psi}_1 = \frac{1+\Gamma^2\psi_2^2}{k^2\sqrt{1+\Gamma^2(\psi_1^2+\psi_2^2)}}\left(\psi_{1.xx}+\psi_{1.yy}\right) - \frac{\Gamma^2\psi_1\psi_2}{k^2\sqrt{1+\Gamma^2(\psi_1^2+\psi_2^2)}}\left(\psi_{2.xx}+\psi_{2.yy}\right)$$
$$+ \frac{2(1+\Gamma^2\psi_2^2)}{k\sqrt{1+\Gamma^2(\psi_1^2+\psi_2^2)}}\left(A_1\psi_{2.x}+A_2\psi_{2.y}\right) + \frac{2\Gamma^2\psi_1\psi_2}{k\sqrt{1+\Gamma^2(\psi_1^2+\psi_2^2)}}\left(A_1\psi_{1.x}+A_2\psi_{1.y}\right) \quad (6)$$
$$+ \frac{\psi_2\sqrt{1+\Gamma^2(\psi_1^2+\psi_2^2)}}{k}\left(A_{1.x}+A_{2.y}\right) - \frac{\psi_1}{\sqrt{1+\Gamma^2(\psi_1^2+\psi_2^2)}}\left(A_1^2+A_2^2\right) + \frac{\psi_1(1-\psi_1^2-\psi_2^2)}{\sqrt{1+\Gamma^2(\psi_1^2+\psi_2^2)}},$$

$$\dot{\psi}_2 = \frac{1+\Gamma^2\psi_1^2}{k^2\sqrt{1+\Gamma^2(\psi_1^2+\psi_2^2)}}\left(\psi_{2.xx}+\psi_{2.yy}\right) - \frac{\Gamma^2\psi_1\psi_2}{k^2\sqrt{1+\Gamma^2(\psi_1^2+\psi_2^2)}}\left(\psi_{1.xx}+\psi_{1.yy}\right)$$
$$- \frac{2(1+\Gamma^2\psi_1^2)}{k\sqrt{1+\Gamma^2(\psi_1^2+\psi_2^2)}}\left(A_1\psi_{1.x}+A_2\psi_{1.y}\right) - \frac{2\Gamma^2\psi_1\psi_2}{k\sqrt{1+\Gamma^2(\psi_1^2+\psi_2^2)}}\left(A_1\psi_{2.x}+A_2\psi_{2.y}\right) \quad (7)$$
$$- \frac{\psi_1\sqrt{1+\Gamma^2(\psi_1^2+\psi_2^2)}}{k}\left(A_{1.x}+A_{2.y}\right) - \frac{\psi_2}{\sqrt{1+\Gamma^2(\psi_1^2+\psi_2^2)}}\left(A_1^2+A_2^2\right) + \frac{\psi_2(1-\psi_1^2-\psi_2^2)}{\sqrt{1+\Gamma^2(\psi_1^2+\psi_2^2)}},$$

$$\sigma\dot{A}_1 = -\frac{1}{k}\left(\psi_2\psi_{1.x}-\psi_1\psi_{2.x}\right) - \left(\psi_1^2+\psi_2^2\right)A_1 + A_{1.yy} - A_{2.xy}, \quad (8)$$

$$\sigma\dot{A}_2 = -\frac{1}{k}\left(\psi_2\psi_{1.y}-\psi_1\psi_{2.y}\right) - \left(\psi_1^2+\psi_2^2\right)A_2 + A_{2.xx} - A_{1.xy}. \quad (9)$$

Here $\psi_1 = \mathrm{Re}\,\psi$, $\psi_2 = \mathrm{Im}\,\psi$, and $A_1$ and $A_2$ are x- and y-components of the vector potential $A$. Also, time derivatives are denoted by dots, e.g., $\dot{A}_1 = \partial A/\partial t$, and spatial derivatives by subscript dots, e.g.: $\psi_{1.y} = \partial\psi_1/\partial y$.

Boundary conditions for the order parameter and the vector potential (components of the current) must be defined. The electrical current itself is expressed as $I = \int J(x,y)dl$, where $J$ is the current density. If $J$ is symmetric relative to the strip's cross-sectional line y=0, then one can confirm (see Ref. [8]) that the magnetic field at the distance $r \geq h/2$ from the horizontal axis of the strip is $B \approx \beta I/r$. Here, $\beta$ is a constant: $\beta \approx 2$, and $h = h(x)$ is the height of the strip. Obviously, $B$ is oppositely directed at the top and bottom edges of the strip, and at these edges, it is orthogonal to the surface of the strip. Taking into account that the $B$–field created by the flow of externally sourced current in thin film strips can be expressed by the vector potential via

$$B = \nabla \times A \mid_z = \frac{\partial A_2}{\partial x} - \frac{\partial A_1}{\partial y} = A_{2.x} - A_{1.y}, \tag{10}$$

the necessary boundary conditions can be formulated: $A_{2.x} - A_{1.y} = B_0 = 2\beta I / h$ on the top boundary line and $A_{2.x} - A_{1.y} = -B_0$ on the bottom boundary line. These boundary conditions at the symmetric (relative to the x-axis) current density distribution guarantee the $I = const$ current flow through the strip. Details regarding the current density distribution near the y-facets of the strip are essential at distances $\sim \xi$ and are not critical to solutions for strips with length greatly exceeding this distance. We have chosen periodic boundary conditions at these edges for the results presented below. Other boundary conditions have also been tried with no essential effect on solutions.

## 4. Finite element approach to the numerical solutions

For our calculations, we used the General Form PDE module of COMSOL version 5.2a [9,10].
The following parameters are used in all of our reported calculations. The Ginzburg-Landau coefficient $\kappa$ was given a value of 4. The total current through the strip was $I = 1.155$, and conductivity $\sigma$ was assigned a value between 1 and 10 (the changes related to the variation of $\sigma$ are found to be nonessential). The superconducting strip was of size $20\lambda_L \times 8\lambda_L$, and the dimensionless electron-phonon relaxation time $\Gamma$ was given a value of 0 or 1 with qualitatively similar outcome.

## 5. Results

The major results of our work were in noticeable influence of the curl-less external vector potential on the dynamics of the vortices in a superconducting strip, Fig 2. Because of gauge invariance, the phase of the $\Psi$–function is known to be coupled with the vector potential. However, in static limit, there are no local consequences of this effect when the vector potential is curl-less. Our equations allow us to track the dynamics of this coupling, as shown in Fig. 2. At $t = 0$ superconductivity and constant vector potential are introduced, and the coupling between the phase and the vector potential is in action (we start with $A = const \neq 0$ and the phase $\theta = 0$ state). This creates a noticeable disturbance: even though we start with the initial condition $\Psi(t = 0) = 1$, which corresponds to the red color in Fig. 2, this state caused by the interaction of the wave-function and the vector potential reduces the modulus of $\Psi$ to zero almost immediately, which corresponds to the blue color in Fig. 2. This corresponds to the first two modeling results shown on the left of the top panel. However, the disturbance quells fairly quickly, and the homogeneous state with $|\Psi| = const$ sets in both the case of $A_{ext} = A_0$ (last entry in the upper row of the upmost panel in Fig. 2) and $A_{ext} = -A_0$ (lower entry in the same panel). Occasionally, even a single-quantum vortex can be created during this initial coupling stage, and will undergo a drift out of the strip.

In the absence of current, a homogeneous superconducting state eventually results independently of the direction of the vector potential, as shown in Fig. 2 (last entries in top panel). We analyzed only the cases $A_{ext} = A_x = A_0$ and $A_{ext} = A_x = -A_0$; these values should be added to the function $A_1$ in Eqs. (6)-(9). As soon as the homogeneous steady state is settled, a DC current is switched on for a short time, which has different consequences depending on how the steady state was achieved: is the external vector potential along or against the current direction. These differences are illustrated in Fig. 2.

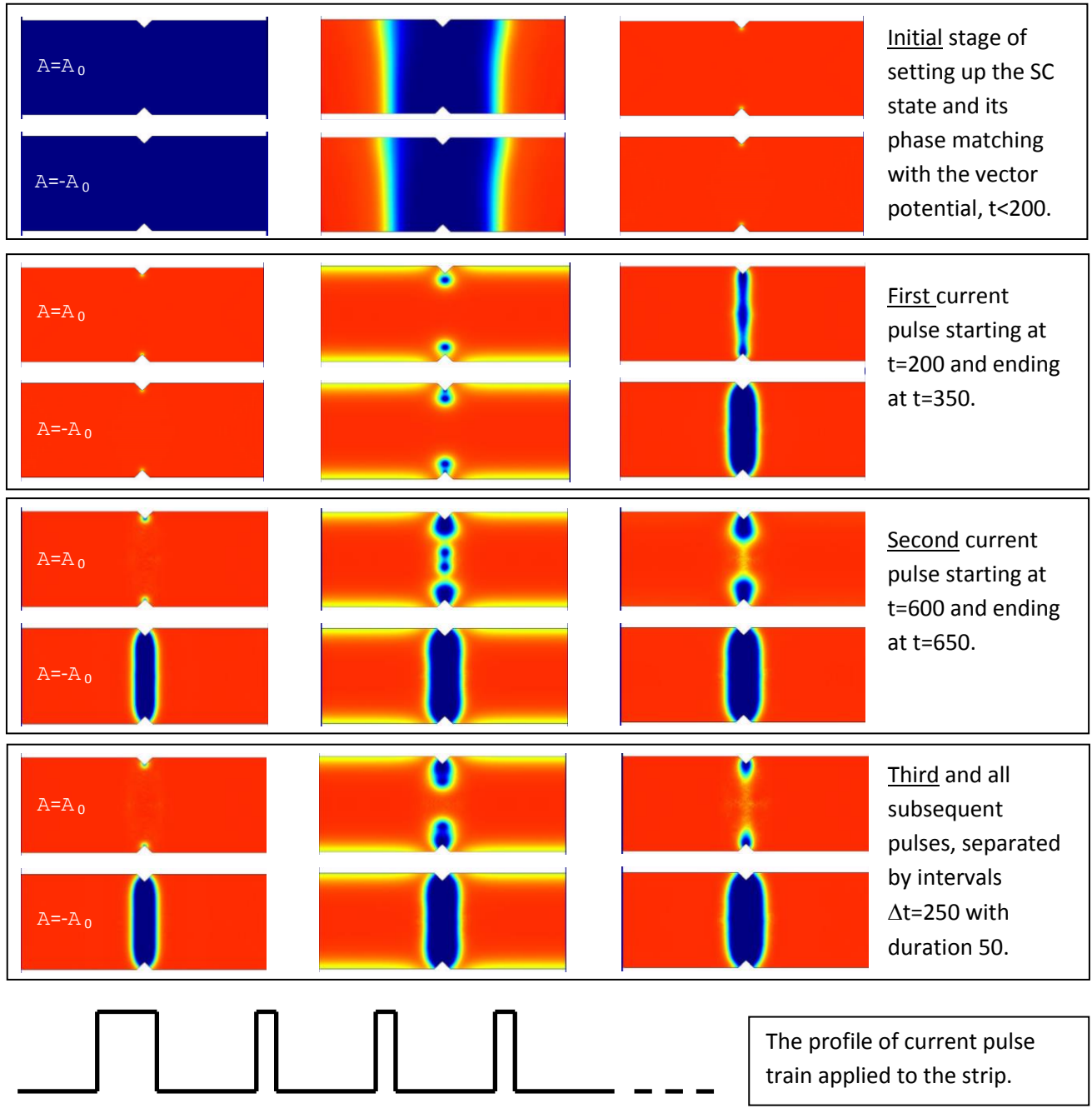

Fig. 2. Results of modeling vortex-antivortex dynamics in current-carrying superconducting strips in presence of the constant external vector potential. Initially (at t=0), the current is absent, and superconducting state is settled similarly and independently on the direction of the vector potential (upper panel). This similarity is broken after the first current pulse, as evident from the second panel. Later, zero resistivity is restored in the case of A=$A_0$, whereas in the case of A=−$A_0$, normal area remains as long as the pulses arrive.

## 6. Discussion

The major consequence which is caused by these differences is the constant presence of resistivity in the case $A_{ext} = -A_0$, while in the case $A_{ext} = A_0$ the resistive state appears for a short time, and is mainly absent. The latter behavior is periodic in time. The resistive state is caused by the normal areas in the middle of the strip. It is interesting to analyze the nature of the normal areas. Intriguingly, the order parameter is fully suppressed (*left panel* in Fig. 3) because of the action of two gigantic vortices of opposite polarity, with magnetic field amplitude shown in *right panel* of Fig. 3.

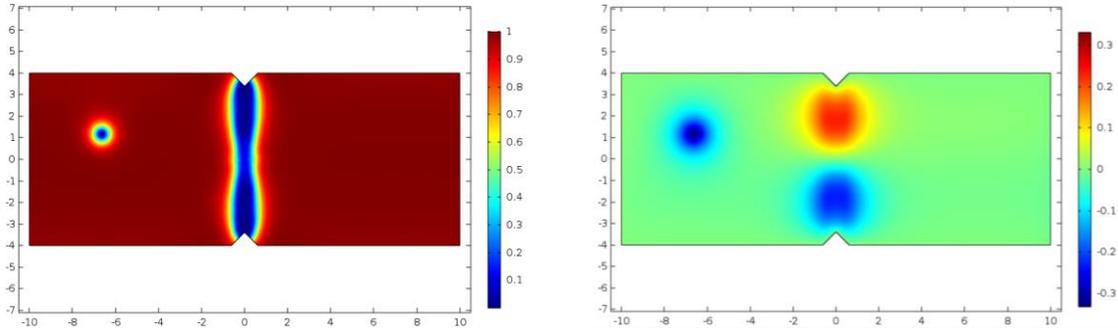

Fig. 3. *Left panel*: the modulus of the order parameter is drastically reduced to zero. Also shown is a separate, single-quantum Abrikosov (anti)vortex; these singular vortices appear episodically at the initial coupling of the phase with the vector potential. The *right panel*: the distribution of magnetic field amplitudes. It is clear that this gigantic vortex-antivortex pair creates the normal area.

Superconducting current circulates around the singular Abrikosov vortex, as well as around the gigantic vortices (red and blue colors in bottom panel correspond to opposite current directions). This state is metastable, and disappears after a much longer time in comparison to the case of a single-quantum vortex-antivortex pair. With the parameters indicated above (in particular, $A_0 = -1.7$), it takes about 8840 units of time for the superconducting connection between the left and right sides of the strip to be restored (*i.e.*, the superfluid flow is restored). That means that if the pulsing current is stopped, then eventually in Fig. 2 the states corresponding to $A_{ext} = -A_0$ and $A_{ext} = A_0$ will result to the same pattern. Contrary, when the pulses are repeatable, this difference shown in Fig. 2 can stay for any required duration.

## 7. Conclusion

We considered dynamic effects caused locally by the curl-less vector potential in small quantum objects (*i.e.*, a current-carrying superconducting strip). In practice, the curl-less vector potential can be caused by very long solenoids or thin magnetic needles with large aspect ratios. These fields are major actors in the case the Aharonov-Bohm effect. Until now, this effect was detected by the electron diffraction effect in a way that guarantees closed electronic wave-function trajectories around the source of the magnetic field. Alternatively, current peculiarities were detected in a closed loop around the AB-potential sources. However, as our dynamic solutions demonstrate, closed looping is not mandatory for detection of the curl-less vector potentials. We predicted classically-detectable effects in quantum objects, such as small current-carrying pieces of superconductors. These objects are demonstrating detectable effects without encompassing remote magnetic fluxes. One may characterize these effects as quasi-local responses of quantum objects to the presence of the curl-less vector potential. This gives us grounds to declare that, in general, the curl-less vector potentials are a detectable field without the involvement of non-locality.

**Acknowledgements**

The authors are grateful to P. Abramian-Barco for the suggestion to test vortex motion in presence of the AB-potential; to Y. Aharonov and J. Tollaksen for enlightening discussions during the recent years; to D. Van Vechten for her interest and support; to J. Przybysz for encouragement. This research was supported by the ONR Grants N00014-15-12095, N00014-16-1-2269, and N00014-16-1-2656.